\documentclass{PoS}
\usepackage{amsmath,amssymb}
\usepackage{caption}
\usepackage{subcaption}

\title{Testing the stochastic LapH method in the twisted mass formulation}

\ShortTitle{Testing the stochastic LapH method in the twisted mass formulation}

\author{\speaker{Christian Jost}, Bastian Knippschild, Carsten Urbach,
  Falk Zimmermann\\
        HISKP (Theory), University of Bonn\\
        E-mail: \email{jost,knippsch,urbach,fzimmerm@hiskp.uni-bonn.de}}

\abstract{
We present first results using the stochastic Laplace-Heaviside (sLapH) method
in the twisted mass formulation.
The calculations are performed on gauge configurations provided by the European Twisted Mass (ETM)
collaboration with 2+1+1 dynamical quark flavours at a single value of the
lattice spacing.
In particular, we compute disconnected contributions to flavour singlet
pseudo-scalar mesons and compare sLapH to standard volume noise methods.
}

\FullConference{31st International Symposium on Lattice Field Theory LATTICE 2013\\
                 July 29 August 3, 2013\\
                 Mainz, Germany}

\newcommand{\gb}{\text{~GB}}
\begin{document}

\section{Introduction}
Lattice QCD provides a non-perturbative method for investigating strong QCD by Markov Chain Monte-Carlo methods in Euclidean space-time.
If one is interested in the hadron spectrum, lattice QCD is a very powerful tool for determining the ground states in a given channel.
Higher excitations or even resonances, however, are much more intricate, because exponentially suppressed contributions must be disentangled from correlator data plagued with statistical noise.

It is well known that correlator matrices are of significant help in such a situation.
With the variational approach~\cite{Michael:1982gb,Luscher:1990ck,Blossier:2009kd} or factorising fit models, excited energy levels can be determined from such a matrix.
Furthermore, the larger the matrix the more energy levels can be extracted with confidence.

However, building a correlator matrix from a large basis of operators is a demanding task, both in computational resources and in storage space.
With these challenges in mind the Laplacian Heaviside smearing (LapH) method~\cite{Peardon:2009gh} and its stochastic version (sLapH)~\cite{Morningstar:2011ka} were invented, leading to remarkable results, see for instance Refs.~\cite{Dudek:2013yja,Moir:2013ub,Dudek:2012xn}.
Its particular strength lies in the possibility to construct a large number of operators from a fixed (but large) number of inversions.

In this proceeding contribution we will investigate sLapH for the case of the Wilson twisted mass formulation of lattice QCD~\cite{Frezzotti:2000nk}.
In particular, we compare sLapH with more conventional methods for the case of the $\eta,\eta'$ mesons, which obtain significant contributions from disconnected diagrams. 

\section{Stochastic LapH Smearing}

The LapH method is a smearing method based on the Laplace operator $\widetilde\Delta$ which is decomposed into its eigensystem 
\begin{equation}
\widetilde\Delta = V_\Delta \Lambda_\Delta V_\Delta^\dagger\,.
\end{equation}
The matrix $V_\Delta$ contains all eigenvectors $v$ and $\Lambda_\Delta$ is a diagonal matrix containing the eigenvalues.
A smearing matrix~$\cal S$
\begin{equation}
  {\cal S} = V_\Delta \Theta \left(\sigma_s^2 + \Lambda_\Delta\right)
  V_\Delta^\dagger = V_s V_s^\dagger\,,
\end{equation}
can be defined, where $\Theta(\cdot)$ is the Heaviside function. 
Therefore, all eigenvalues larger than a cutoff $\sigma_s^2$ are neglected and the matrix $V_s$ consists only of the eigenvectors belonging to the eigenvalues smaller than $\sigma_s^2$.
The number of eigenvalues needed depends on the spatial volume of the lattice and the parameter $\sigma_s^2$.
It was found that $\sigma_s^2=0.33$ is optimal \cite{Morningstar:2011ka} for excited state suppression independent of the operator.
On a $48^3\times96$ lattice, for example, this amounts to $900$ eigenvectors per timeslice.
In the same reference it was shown that the shape of the source is approximately Gaussian. 

A quark line $\cal Q$ can be written in terms of the smearing matrix
\begin{equation}\label{eqn:quarkline}
  {\cal Q} = D^{(j)}{\cal S}\Omega^{-1}{\cal S}D^{(k)\dagger} =
  D^{(j)} V_s\ (V_s^\dagger \Omega^{-1} V_s) 
  \ V_s^\dagger D^{(k)\dagger}\,,
\end{equation} 
where $D^{(j)}$ is a covariant derivative.
${\cal P} = V_s^\dagger \Omega^{-1} V_s$ is called perambulator, $\Omega = \gamma_4 M$, $\gamma_4$ is a Dirac matrix, and $M$ being the lattice Dirac operator.
The computation of $\Omega^{-1}v$ must be performed for each eigenvector $v$, which is a significant drawback of this method, in particular for large lattice volumes.

\subsection{Stochastic LapH}

In order to overcome this drawback the stochastic LapH method was introduced in Ref.~\cite{Morningstar:2011ka}.
Random noise vectors $\rho$ which live in the time, Dirac- and Laplace eigenspace should reduce the computational cost and not change the result significantly if the following
relations hold
\begin{equation}
  E(\rho_i)=0\quad \text{ and }\quad E(\rho_i\rho_j^\ast)=\delta_{ij}\,,
\end{equation}
where $E(\cdot)$ denotes the expectation value.
The propagator $\Omega^{-1}$ can then be calculated approximately by solving
\begin{equation} 
  \Omega X^r = \rho^r
\end{equation}
for $N_R$ different random vectors $\rho^r$.
Dilution can be used to enhance the signal-to-noise ratio.
The improvement factor is proportional to $\frac{1}{N_d\sqrt{N_R}}$, where $N_d$ is the number of dilution vectors.
Therefore, the signal-to-noise ratio ameliorates proportional to the number of inversions done.
The dilution factorises into three subspaces, time, Dirac- and eigenspace and thus different dilution scheme can be applied in each subspace.
In Ref.~\cite{Morningstar:2011ka} four basic dilution schemes are discussed
\begin{equation}
  \begin{array}{lll}
    P^{(b)}_{ij} = \delta_{ij},               & b=0,
    & \mbox{(no dilution)} \\
    P^{(b)}_{ij} = \delta_{ij}\ \delta_{bi},  & b=0,\dots,N-1, 
    & \mbox{(full dilution)}\\
    P^{(b)}_{ij} = \delta_{ij}\ \delta_{b,\, \lfloor Ji/N\rfloor}, 
    & b=0,\dots,J-1, & \mbox{(block-$J$)}\\
    P^{(b)}_{ij} = \delta_{ij}\ \delta_{b,\, i\bmod J}, & b=0,\dots,J-1, 
    & \mbox{(interlace-$J$)}\,,\nonumber
  \end{array}
\end{equation}
where $N$ is the size of the subspace (time, Dirac- and eigenspace) and $N/J$ is an integer.
$N_d$ is 1 for no dilution, $N$ for full dilution and $J$ for interlace or block dilution.
Another possibility for a dilution scheme would be to combine the interlace and the block dilution schemes.

Using random noise vectors and a dilution scheme, one arrives at a similar expression for the stochastic quark lines as in eq.~\ref{eqn:quarkline}
\begin{equation}
  {\cal Q} = \textstyle\sum_b E\Bigl( \! D^{(j)} {\cal S} \Omega^{-1} V_sP^{(b)}\rho
  \, (D^{(k)} V_s P^{(b)}  \rho)^\dagger \!\Bigr)\,.
\end{equation}
Here, the first part, $\phi= D^{(j)} {\cal S} \Omega^{-1} V_sP^{(b)}\rho$, acts as a quark sink and the second part, $\eta=D^{(k)} V_s P^{(b)} \rho$ as a quark source.
A stochastic perambulator ${\cal P}_s$ can be defined
\begin{equation}
  {\cal P}_s = V_s \Omega^{-1} V_sP^{(b)}\rho\,.
\end{equation}
Its computation is the computationally most expensive part of this method. 
However, the number of inversions which need to be performed only depends on the number of random vectors and the dilution scheme. The volume dependence of the non-stochastic LapH method has almost disappeared and appears only very mildly in certain dilution schemes if at all.
%For any dilution combination that does not use full dilution in time and LapH eigenspace,
%the number of inversions is independent of the lattice size and the number of eigenvectors.
This might be useful for larger lattices with many eigenvectors.

To illustrate the problem of computational cost and storage, the number on inversions and the storage space needed for a $48^3\times96$ lattice is calculated.
900 eigenvectors per timeslice are needed and hence the matrix $V_s$ would need roughly $427\gb$ of storage space.
To calculate the full perambulator $\cal P$, $900\cdot96\cdot 4=345600$ inversions would have to be performed and the final perambulator would need approximately $1780\gb$.
The computational cost of the inversions and the storage requirements make it unfeasible to use this method for larger lattices at the moment. 

Using the stochastic ansatz reduces the computational cost significantly.
With full dilution in time and in Diracspace and an interlace-8 dilution in eigenspace (TF,DF,LI8), the number of inversions would reduced to 3072, while the storage space would reduce to $15.8\gb$.
Usually, it is not necessary to use full dilution in time, the computational cost could be reduced further.
Using an interlace-16 scheme in time (TI16,DF,LI8) would result in 512 inversions and a storage size of only $2.6\gb$.
The latter scheme is entirely independent of the lattice size.

%\section{Implementation}

%\textbf{BK: Das interessiert doch niemanden. Ich wrde den teil komplett rausnehmen und nur den tmLQCD code kurz in der nchsten section erwhnen.}
%As a code basis we used the tmLQCD code~\cite{Jansen:2009xp}.
%The calculations of the eigensystem were performed with the tmLQCD code using an implementation of the Jacobi-Davidson algorithm.
%At the same time we are testing PETSc\footnote{Portable, Extensible Toolkit for Scientific Computation (PETSc), linear algebra library} and SLEPc~\cite{Hernandez:2005:SSF}\footnote{Scalable Library for Eigenvalue Problem Computations (SLEPc)} to have several different eigensystem solvers and preconditioners available.
%Moreover both libraries are already parallelized with MPI.
%We used a mixed precision CG algorithm implemented for GPUs with even/odd preconditioning to compute the propagators.

%The contraction code is based on the EIGEN~\cite{eigenweb} and BOOST~\cite{boost_lib} libraries.
%We implemented the ILDG data format for reading gauge and propagators files.
%The code computes all Dirac matrix combinations at source and sink.
%Displacement and momentum operators are not yet implemented.
%The code is already parallelized with OpenMP, MPI support should be coming soon. 

\section{Numerical Experiments}

\subsection{Simulation parameters}

Our simulations were performed on the $A60.24$ ensemble with $N_f=2+1+1$ dynamical Wilson twisted mass fermions as produced by the European Twisted Mass collaboration~\cite{Baron:2010bv,Baron:2010th,Baron:2011sf}. 
The volume of this lattice is $V=24^3\cdot 48$ with a lattice spacing of $a=0.086\text{ fm}$, and a pion mass of about 390 MeV.
In total, 314 gauge configurations were used.
%More details on the ensemble can be found in the reference mentioned before.
For strange and charm quarks the so-called Osterwalder-Seiler mixed action~\cite{Frezzotti:2004wz} with $a\mu_s=0.02322$ and $a\mu_c=0.27678$ (c.\ f. \cite{Bonn:2013OS}) was used.
More details on this approach can be found in a contribution to this conference~\cite{Bonn:2013OS}.
The inversions and the computation of the eigenvectors were performed
with the tmLQCD software suite~\cite{Jansen:2009xp}.

The Euclidean correlation functions for the $\eta,\eta'$ system is given by
\begin{equation}
  \label{eq:correlations}
  \mathcal{C}(t)_{qq'} =
  \langle\mathcal{O}_q(t'+t)\mathcal{O}_{q'}(t')\rangle\,,\quad
  q,q'\in{\ell,s,c}\,, 
\end{equation}
where the operators $\mathcal{O}_\ell = (\bar ui\gamma_5 u + \bar d i\gamma_5 d)/\sqrt{2}$, $\mathcal{O}_s = \bar s i\gamma_5 s$ and $\mathcal{O}_c = \bar c i\gamma_5 c$ were used.
For this operator basis the resulting matrix $\mathcal{C}$ is a $3\times3$ matrix which is diagonalised by solving the Generalised Eigenvalue Problem. The effective masses
\begin{equation}
aM^{(n)} = -\log(\lambda^{(n)}(t,t_0)/(\lambda^{(n)}(t+1,t_0))
\end{equation}
are computed from the eigenvalues $\lambda^{(n)}(t,t_0)$.
In this ansatz the $\eta$ meson corresponds to the ground state ($n=0$) and the $\eta^\prime$ meson to the first excited state ($n=1$).
The second excited state ($n=2$) corresponds to the $\eta_c$ meson but cannot be resolved due to the small correlation matrix.
In addition, we investigate the disconnected contribution to the neutral pion, corresponding to the operator ${\cal O}_{\pi^0} = (\bar ui\gamma_5 u - \bar d i\gamma_5 d)/\sqrt{2}$.

For the calculation of the eigenspace of the Laplace operator, two
iterations of 3D HEX~\cite{Capitani:2006ni} smearing with parameters
$(0.76,0.95)$ were applied to the spatial gauge fields. 
On each timeslice we calculated 120 eigenvectors and we used the previously introduced dilution scheme (TF, SF, LI8).
%Our light quark simulations were performed with two dilution schemes: (TF, SF, LI8) and (TI16, SF, LI8).
%The simulation of the heavy quarks were only done with full dilution in time.

The sLapH method is tested against a standard approach of computing correlation functions.
In this standard approach, connected diagrams were computed by using stochastic timeslice sources and the so called ``one-end-trick''~\cite{Boucaud:2008xu}.
%For this we rely on stochastic sources in order to calculate connected diagrams via the so called ``one-end-trick''~\cite{Boucaud:2008xu}.
Disconnected diagrams which need an all-to-all approach were sampled using 24 stochastic volume sources with complex Gaussian noise.
Two different methods were used to improve the signal-to-noise ratio. The first one is a hopping parameter variance reduction scheme, see Ref.~\cite{Boucaud:2008xu}, which will be denoted with sGVS.
The second one is only available for the $\eta,\eta'$ system in the twisted mass formulation and it relies on the identity~\cite{Jansen:2008wv}
\begin{equation}\label{eq:vv}
  D^{-1}_u-D^{-1}_d = -2i \mu_l D^{-1}_d \gamma_5 D^{-1}_u\,,
\end{equation}
where $D_{u,d}$ are the up and down lattice Dirac operators and $\mu_l$ is the twisted mass parameter corresponding to the light quarks.
It can be applied for strange and charm quarks as well by introducing
a doublet of strange (charm) quarks, see Ref.~\cite{Bonn:2013OS} for
details. 
This noise reduction will be denoted by improved sGVS.

The efficiency of sLapH as a smearing method to suppress excited state contribution can be tested by comparing the effective mass plateaus computed from sLapH, sGVS and improved sGVS. 
In the latter two methods only local operators were used and the size of the correlation matrix was 3x3. 
Hence, a direct comparison was possible and a notable reduction of excited state contributions in case of sLapH should be observable.

\subsection{Correlation functions}

\begin{figure}[t]
  \centering
  \begin{subfigure}[b]{.47\textwidth}
    \includegraphics[width=\textwidth]{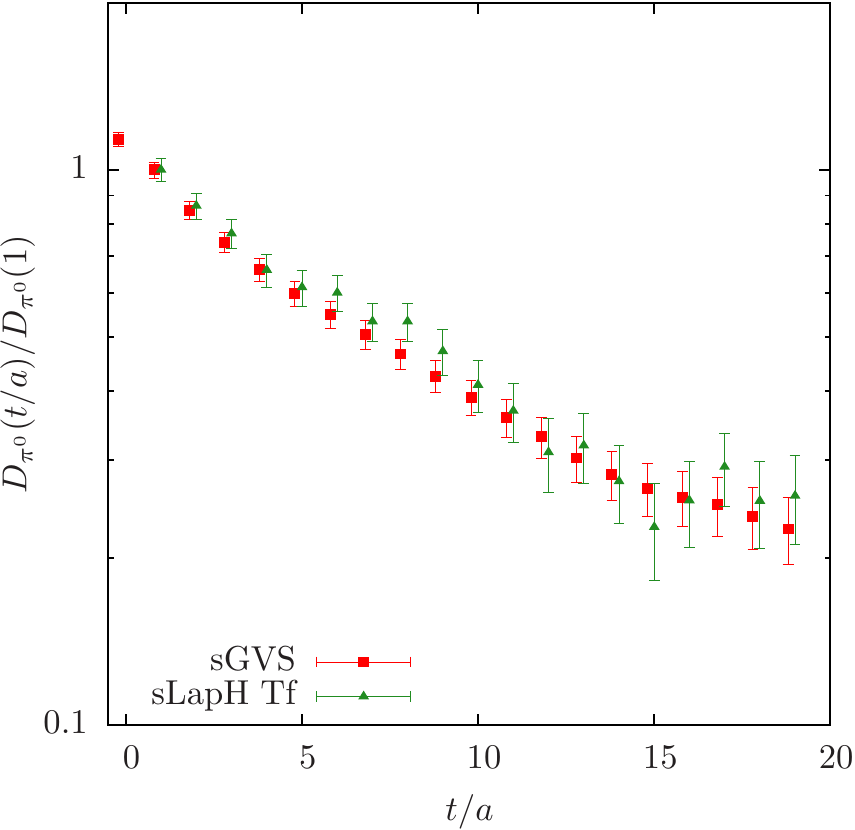}
  \end{subfigure}
  \quad
  \begin{subfigure}[b]{.49\textwidth}
    \includegraphics[width=\textwidth]{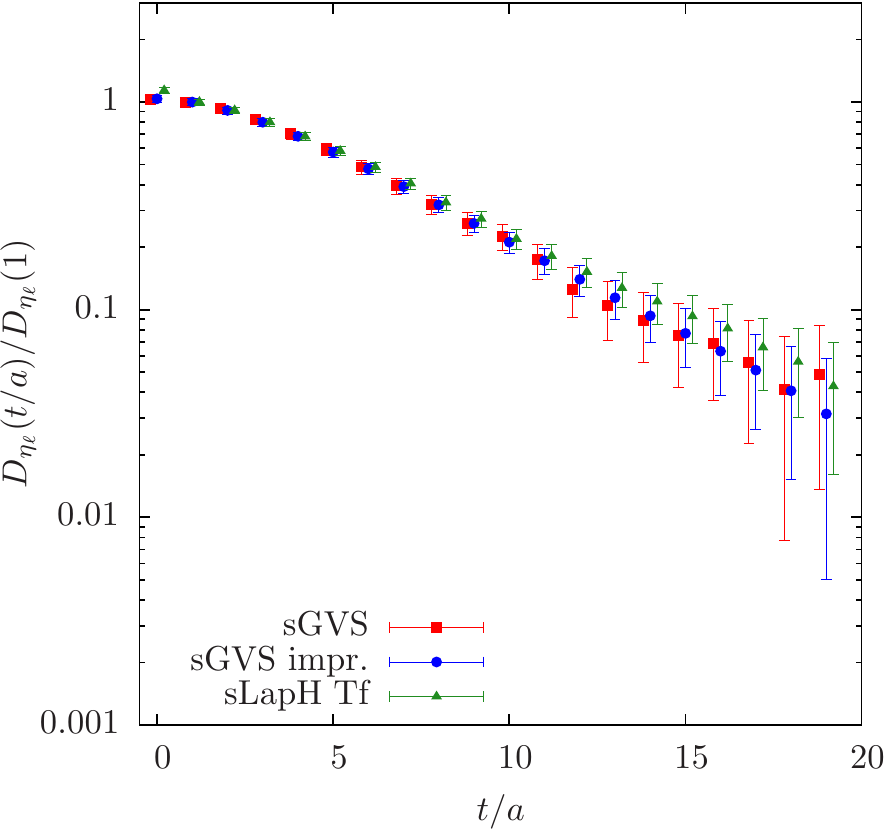}
  \end{subfigure}
  \caption{Disconnected contribution to the $\pi^0$ (left) and the $\eta_\ell$ (right) correlation functions are shown.
    Both are normalised to $t/a=1$. The points are slightly displaced for better legibility.}
  \label{fig:disconn}
\end{figure}

The disconnected parts of the correlation functions for the $\pi^0$ and the $\eta_\ell$ are shown in Figure~\ref{fig:disconn}. 
They are normalised to $D(1)$ due to an unknown normalisation factor in sLapH.
The $\eta_\ell$ contains only the light quark contributions to the $\eta$ and $\eta^\prime$ mesons.
Within errors, the results obtained from different methods for these observables do agree quite well.
While the errors for the disconnected part of the $\pi^0$ are compatible in between sLapH and sGVS, sLapH and improved sGVS turn out to outperform sGVS for the disconnected part of the  $\eta_\ell$.
sLapH and improved sGVS give roughly equal error estimates with a slight advantage for sLapH.
However, the computation cost is a factor $24/512$ in favour of improved sGVS.

\subsection{Effective Masses}

\begin{figure}[t]
  \centering
  \begin{subfigure}[b]{.48\textwidth}
    \includegraphics[width=\textwidth]{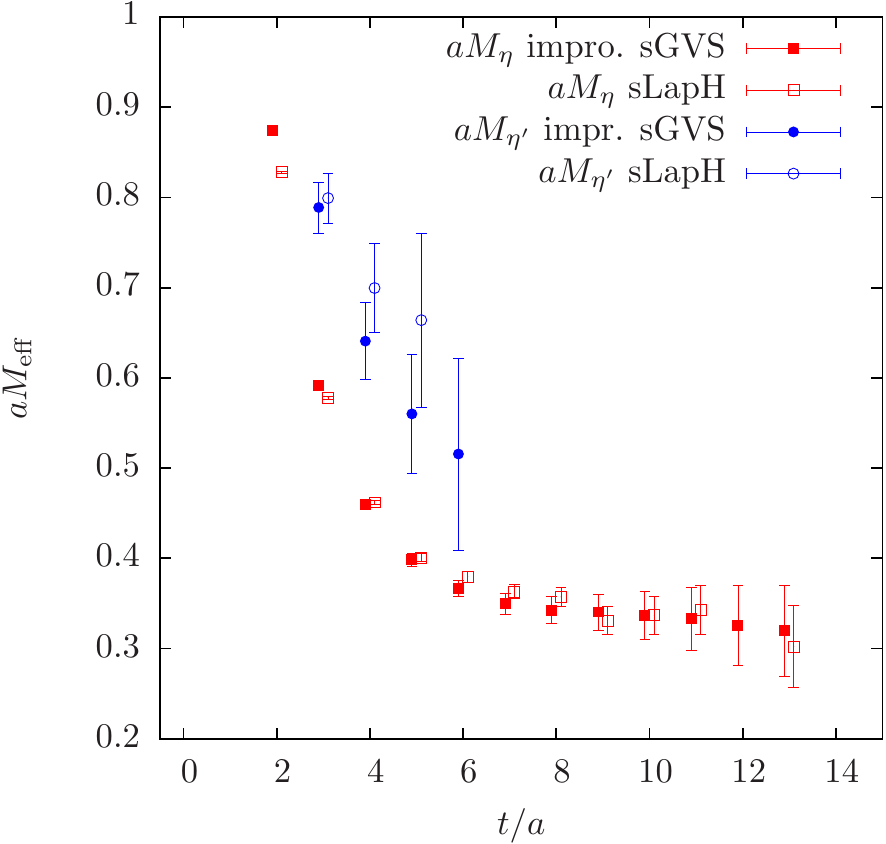}
  \end{subfigure}
  \quad
  \begin{subfigure}[b]{.48\textwidth}
    \includegraphics[width=\textwidth]{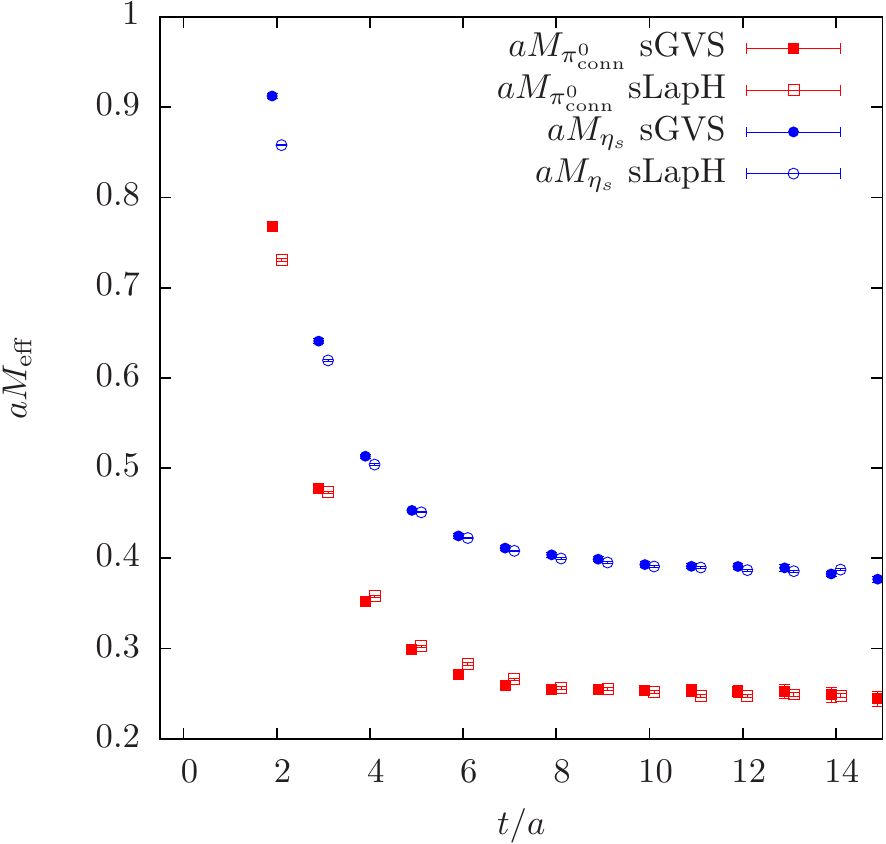}
  \end{subfigure}
  \caption{(left) Effective mass plot of the $\eta$ (squares) and $\eta^\prime$ (circles) using the sGVS method (full symbols) and the sLapH method (empty symbols).
    (right) The same as (left), but with connected contributions only.
    The points are slightly displaced for better legibility.} 
  \label{fig:mass}
\end{figure}

In the left panel of Figure~\ref{fig:mass} the effective masses of the $\eta$ (filled symbols) and the $\eta'$ (open symbols) are shown for the sGVS and sLapH method.
General agreement within errors can be observed.
However, the number of gauge configurations  and the size of the correlation matrix are too small to obtain a good signal for the $\eta'$ state.

Unfortunately, sLapH seems not to suppress excited states very much, only at $t/a=2,3$ a small effect can be observed. 
This can be seen even better by looking at the connected correlators only, corresponding to a connected neutral pion and the so called $\eta_s$ state, a pion made out of strange quarks.
The effective masses of these states are shown in the right panel of figure~\ref{fig:mass}. 
The poor suppression of excited state contributions might be due to poorly chosen parameters for the link smearing used in the Laplace operator. 
The eigenvalues of the Laplace operator are squeezed quite strongly for our choice of smearing parameters compared to unsmeared gauge links.
This results in a more shallow increase of the eigenvalues. In conclusion, much more eigenvectors would be needed to achieve a good reduction of excited state contributions. 
Well chosen link smearing parameters should have the very opposite effect and stretch the spectrum of the Laplace operator. 
Hence, less vectors would be needed to gain a good reduction.

\section{Summary}

In this article we presented a first test of the sLapH method in comparison to standard stochastic methods in the twisted mass formulation for the $\eta$, $\eta'$ mesons. 
The results for the meson masses extracted from the different methods do agree within errors but with the sLapH method being roughly 20 times more expensive. 
However, the higher expense, which comes mainly from the increased number of inversions, can be put into perspective by noting that the inversions, which can be easily stored in form of perambulators on a hard drive, can be reused for other observables. 
In addition, the contractions to build correlation functions are cheaper for the sLapH method compared to standard methods which would become more relevant for a larger operator basis. 

As a surprise we did not observe that sLapH suppresses excited states efficiently. In fact, in a comparison to a local-local correlator matrix we see only marginal improvements. 
This is most probably due to a bad tuning of link smearing in the Laplace operator. 
The explicit impact of the link smearing and the impact of the lattice spacing on the efficiency of sLapH are under investigation at the moment.

Other physical channels will be studied as well. 
sLapH clearly still has the advantage of providing the possibility to compute many operators for a fixed number of inversions.

We thank all members of ETMC for the most enjoyable collaboration.
This work is supported in part by DFG and NSFC (CRC 110).

\bibliographystyle{h-physrev5}
\bibliography{bibliography}

\end{document}